\documentclass[a4paper]{article}
\usepackage{graphicx}
\usepackage{amsmath} 
\usepackage{graphicx} % Required for inserting images
\usepackage{amsmath}
\usepackage{booktabs}
\usepackage{physics}
\usepackage{color,soul}
\usepackage{tikz}
\usepackage[margin=1.5cm]{geometry}
\usepackage{graphicx,subcaption,lipsum}
\title{Photon Blockade in Cavity Magnomechanical Systems using Phase-Controlled Feedback}
\author{Mounes Eslami\thanks{Email: mounes.eslami@gmail.com} $\,$, 
Kurosh Javidan\thanks{Email: javidan@um.ac.ir} \\ \\
\small{Department of Physics, Faculty of Science, Ferdowsi University of Mashhad,}\\
\small{9177948974  Mashhad, Iran} }
\date{}
\begin{document}

	\flushbottom
	\maketitle
	\thispagestyle{empty}
	\section*{Abstract}
		In this paper, we optimize photon blockade in a cavity magnomechanical system using feedback by introducing optimized values for the phase and magnetic field coupling strength at each drive frequency. It is shown that the computed values significantly reduce the photon second-order correlation function in the dynamic Schrodinger equation. The Radau method, an implicit Runge-Kutta method, has been employed, which provides more accurate results. Furthermore, we demonstrate that a frequency detuning between the magnon and photon can result in deep values of photon blockade. Utilizing these optimized parameters outperforms scenarios that rely on constant, non-optimized values. This approach provides strong potential for applications in quantum sensing and quantum computation.

	\section*{Introduction}
	Cavity quantum electrodynamics (QED) \cite{walther2006cavity} and optomechanical cavities represent two of the most versatile platforms for studying light-matter interactions at the quantum level. Cavity QED explores the interaction of photons with atoms or artificial atoms confined in optical or microwave resonators, leading to breakthroughs in single-photon generation \cite{kuhn2010cavity, mucke2013generation}, quantum memories \cite{reagor2016quantum}, photonic quantum gates \cite{zheng2013waveguide}, and other applications \cite{najer2019gated}. Similarly, optomechanical systems have enabled remarkable advances in the control of mechanical motion at the quantum limit. Hybrid systems (the integration of these two platforms) allow for the coherent exchange of information between photonic, atomic, and mechanical degrees of freedom, creating opportunities for new applications in quantum information processing \cite{stannigel2011optomechanical, rogers2014hybrid}, high accuracy and fine measurements and many more technological applications\cite{motazedifard2021ultraprecision, kurizki2015quantum}. 
	
	Feedback, traditionally used in many microwave and optical systems, can be employed in quantum systems as well. Feedback allows the system to refine its parameters, minimizing errors and improving outcomes. In the reference  \cite{amazioug2023enhancement}, feedback loop receives the system's output and uses it to control characteristics of input laser. This enhances the entanglement between magnon, photon and phonon in cavity magnomechanics and can be applied in quantum information. The same improved results for both entanglement and quantum steering are predicted by \cite{linquantum}, the system maintains quantum correlations at higher temperatures (up to 0.43 K) due to feedback. 
	
	Photon blockade is a nonclassical phenomenon that arises when a quantum system permits the absorption or emission of a single photon while blocking subsequent photons. Conventional photon blockade, which was first observed in \cite{birnbaum2005photon} where the excitation of the atom- cavity system caused by a single photon blocks the transmission of the second photon. While the unconventional photon blockade relies on quantum interference between multiple pathways for photon excitation, even in systems with weak nonlinearity \cite{flayac2017unconventional}. The concept of blockade has been extended beyond photons to include quasiparticles such as magnons, which was first studied in \cite{liu2019magnon}. Also magnon blockade can occur using Kerr effect \cite{hou2024magnon} and/or through magnon squeezing \cite{amazioug2024achieving}. 
	
	Here, we study the effect of feedback on photon blockade. This phenomenon is controlled using an optimum value for the phase of the feedback which is theoretically obtained in the steady state situation and numerically calculated in dynamical conditions. We show that this approach is able to successfully control and improve the photon blockade. In terms of simulation, we use the Radau method which is an implicit Runge-Kutta method \cite{hairer1999stiff} to solve Shrodinger's equation.

	\section*{The model}
	\begin{figure}[b!]
		\centering
		\includegraphics[scale=0.8]{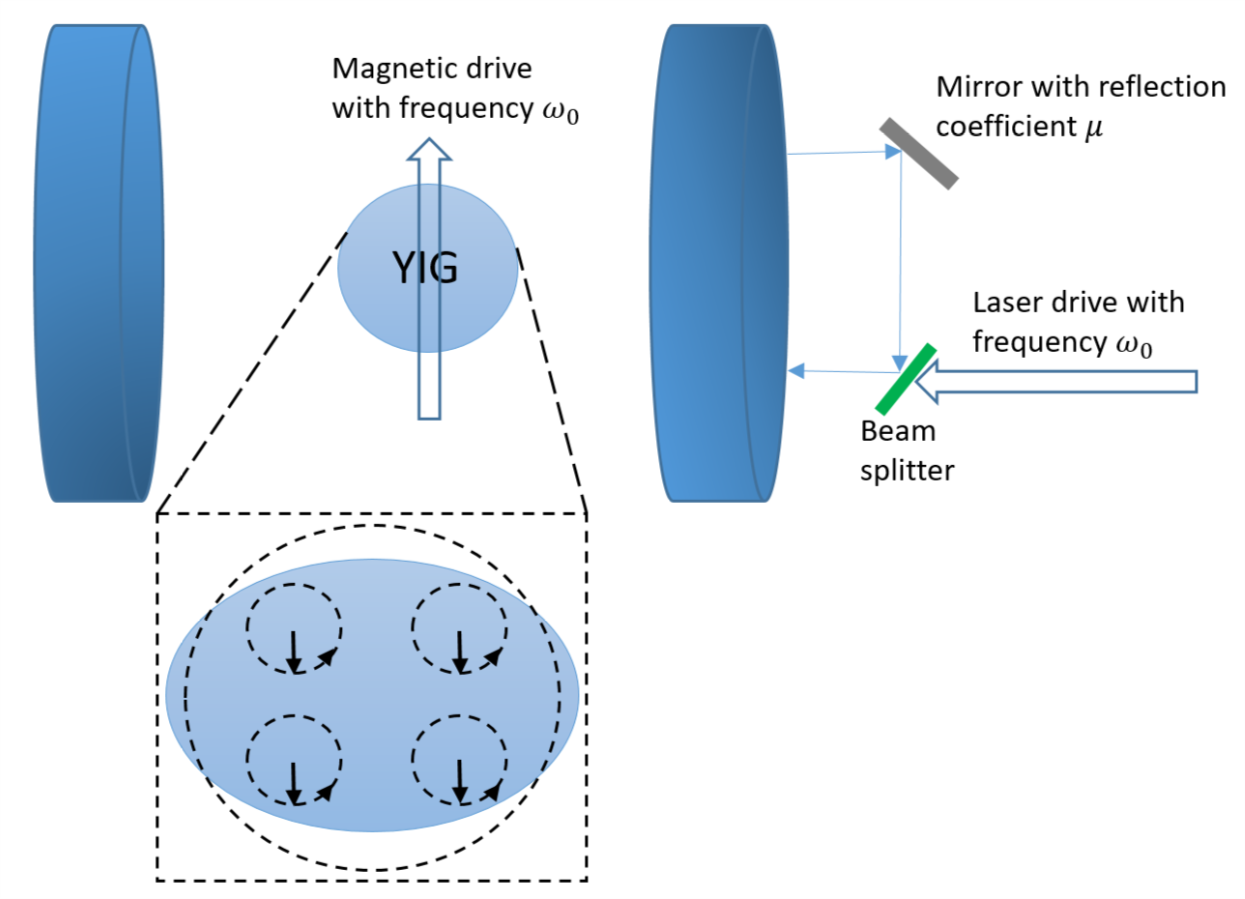} 
		\caption{Schematic diagram of the system. } 
		\label{systemoutline.png}
	\end{figure}
	Fig. \ref{systemoutline.png} shows a general schematic of a cavity magno-mechanical system including optical feedback.
	The cavity contains a YIG with Kittel magnonic mode and also a mechanical mode. The YIG has been fixed in the antinode of the cavity field while being biased by an external drive field. The Hamiltonian is composed of \cite{amazioug2023enhancement}:
	
	\begin{equation} 
	H = H_{free}+H_{md}+H_{mc}+H_{dm}+H_{dc} 
	\end{equation}
	the free Hamiltonian of the system is :
	\begin{equation} 
	H_{free} = \hbar \omega_c c^\dagger c + \hbar \omega_m m^\dagger m + \hbar\omega_d b^\dagger b 
	\end{equation}
	where, $\omega_c$, $\omega_m$, and $\omega_d$ represent the frequencies of the cavity mode, the magnon mode, and the mechanical mode, respectively, while the operators $c^\dagger (c)$, $m^\dagger (m)$, and $b^\dagger (b)$ are the creation (annihilation) operators for the respective modes. The interaction between the magnon and  the mechanical mode is expressed as :
	\begin{equation} 
	\label{H_md}
	H_{md} = \hbar g_{md} m^\dagger m (b + b^\dagger) 
	\end{equation}
	where the coupling strength is $g_{md}$. The following term of the Hamiltonian describes photon-magnon coupling :
	\begin{equation} 
	\label{H_mc}
	H_{mc} = \hbar g_{mc} (c m^\dagger+c^\dagger m) 
	\end{equation}
	which allows energy exchange between the electromagnetic mode in the cavity and the spin wave (magnon) mode in the magnetic material via the coupling strength $g_{mc}$. The Hamiltonian term related to the drive of the magnon mode is :
	\begin{equation} 
	\label{H_dm}
	H_{dm} = i\hbar E(m^\dagger e^{-i\omega_0 t}-me^{i\omega_0 t}) 
	\end{equation}
	where an external time-dependent microwave field (pump) with frequency $\omega_0$ drives the magnon mode, and $E$ is the coupling strength of this field. The same drive frequency is used in the optical feedback mode, represented by :
	\begin{equation} 
	\label{H_dc}
	H_{dc} = \hbar \Omega \mu (c^\dagger e^{-i\omega_0 t} e^{i\phi} + c e^{i\omega_0 t} e^{-i\phi}) 
	\end{equation}
	where the coupling strength is denoted with $\Omega \mu$ and the phase $\phi$ allows for control over the relative phase of the driving field.
	
	The pump (drive) frequency ($\omega_0$) is the same for both magnon and input laser beam, Which is a common assumption used in \cite{amazioug2023feedback, zhao2021phase} and other studies as well. The transformation of a Hamiltonian 
	H under a time-dependent unitary transformation U(t) can be written as: 
	\begin{equation}  
	H^{'} = U H U^\dagger - i U \frac{\partial U^\dagger}{\partial t}
	\end{equation}
	Using the unitary transformation mentioned in \cite{parai2023unconventional}, we have:
	\begin{equation}  
	U = exp(-i \omega_0 t (c^\dagger c + m^\dagger m) )
	\end{equation}
	So, the total Hamiltonian is transformed into : 
	\begin{equation}
	\label{Hamiltonian_without_loss}
	\begin{split}
	H^{'} = & \hbar (\omega_c - \omega_0 ) c^\dagger c + \hbar (\omega_m - \omega_0) m^\dagger m + \hbar\omega_d b^\dagger b  \\
	+ & \hbar g_{md} m^\dagger m (b + b^\dagger) + \hbar g_{mc} (c m^\dagger + c^\dagger m)  \\
	+ & i\hbar E(m^\dagger  - m) +  \hbar \Omega \mu (c^\dagger  e^{i\phi} + c  e^{-i\phi})
	\end{split}
	\end{equation}
	Considering the loss in the photon, magnon and phonon qubit decay ($\kappa_c, \kappa_m, \kappa_d$), and using the non-Hermitian approach, the Hamiltonian changes to: 
	\begin{equation}
	\begin{split}
	H^{'} = & \hbar (\Delta_c ) c^\dagger c + \hbar (\Delta_m) m^\dagger m + \hbar\Delta_d b^\dagger b + \\
	& \hbar g_{md} m^\dagger m (b + b^\dagger) + \hbar g_{mc} (c m^\dagger + c^\dagger m) + \\
	& i\hbar E(m^\dagger  - m) +  \hbar \Omega \mu (c^\dagger  e^{i\phi} + c  e^{-i\phi})
	\end{split}
	\end{equation}
	where:
	\begin{equation}
	\begin{split}
	\Delta_c &= \omega_c - \omega_0 -i \frac{\kappa_c}{2} \\
	\Delta_m &= \omega_m - \omega_0 -i \frac{\kappa_m}{2} \\
	\Delta_d &= \omega_d  -i \frac{\kappa_d}{2}
	\end{split}
	\end{equation}
	
	In this paper we consider more than two photons (two magnons) truncation, thus we can use the following wave function: 
	\begin{equation}
	\begin{split}
	\ket{\psi} = 
	& C_{000} \ket{000} + C_{100} \ket{100} + C_{010} \ket{010}  + C_{001} \ket{001}\\
	& + C_{110} \ket{110}  + C_{101} \ket{101}  + C_{011} \ket{011} \\
	& + C_{200} \ket{200} + C_{020} \ket{020} + C_{002} \ket{002}
	\end{split}
	\end{equation}
	Using Shrodinger's equation we will have the following equations that determine the time evolution of each coefficient in the wave function: 
	\begin{equation} \label{eq:exact}
	\begin{split}
	i \hbar \frac{\partial}{\partial t} C_{000} &= \hbar \Omega \mu e^{-i \phi} C_{100} + (-i \hbar E) C_{010}, \\
	i \hbar \frac{\partial}{\partial t} C_{100} &= \hbar \Omega \mu e^{i \phi} C_{000} + (\hbar \Delta_c) C_{100} + (\hbar g_{mc}) C_{010} \\
	&\quad + (-i \hbar E) C_{110} + (\hbar \Omega \mu e^{-i \phi} \sqrt{2}) C_{200}, \\
	i \hbar \frac{\partial}{\partial t} C_{010} &= (i \hbar E) C_{000} + (\hbar g_{mc}) C_{100} + (\hbar \Delta_m) C_{010} \\
	&\quad + (\hbar \Omega \mu e^{-i \phi}) C_{110} + (\hbar g_{md}) C_{011} + (-i \hbar E \sqrt{2}) C_{020}, \\
	i \hbar \frac{\partial}{\partial t} C_{001} &= \hbar \Delta_d C_{001} + \hbar \Omega \mu e^{-i \phi} C_{101} + (-i \hbar E) C_{011}, \\
	i \hbar \frac{\partial}{\partial t} C_{110} &= (i \hbar E) C_{100} + (\hbar \Omega \mu e^{i \phi}) C_{010} + (\hbar \Delta_c) C_{110} \\
	&\quad + (\hbar \Delta_m) C_{110} + \hbar g_{mc} \sqrt{2} \, C_{200} + \hbar g_{mc} \sqrt{2} \, C_{020}, \\
	i \hbar \frac{\partial}{\partial t} C_{101} &= \hbar \Omega \mu e^{i \phi} C_{001} + (\hbar \Delta_c) C_{101} + (\hbar \Delta_d) C_{101} + (\hbar g_{mc}) C_{011}, \\
	i \hbar \frac{\partial}{\partial t} C_{011} &= (\hbar g_{md}) C_{010} + i \hbar E C_{001} + (\hbar g_{mc}) C_{101} + (\hbar \Delta_m) C_{011} \\
	&\quad + (\hbar \Delta_d) C_{011}, \\
	i \hbar \frac{\partial}{\partial t} C_{200} &= \hbar \Omega \mu e^{i \phi} \sqrt{2} \, C_{100} + \hbar g_{mc} \sqrt{2} \, C_{110} + (2 \hbar \Delta_c) C_{200}, \\
	i \hbar \frac{\partial}{\partial t} C_{020} &= i \hbar E \sqrt{2} \, C_{010} + \hbar g_{mc} \sqrt{2} \, C_{110} + (2 \hbar \Delta_m) C_{020}\\
	i \hbar \frac{\partial}{\partial t} C_{002} &= 2 \hbar \Delta_d C_{002}.
	\end{split}
	\end{equation}
	
	In the steady state situation, the time derivative of the coefficients is zero. We also use the assumption that $\abs{C_{000}} \gg \abs{C_{100}}, \abs{C_{010}}, \abs{C_{001}} \gg \abs{C_{110}}, \abs{C_{101}}, \abs{C_{011}}, \abs{C_{200}}, \abs{C_{020}}, \abs{C_{002}}$, so we will have: 
	\begin{equation} \label{eq:steady_state}
	\begin{split}
	%0 &=  \Omega \mu e^{-i \phi} C_{100} + (-i  E) C_{010}, \\
	0 &=  \Omega \mu e^{i \phi} C_{000} + ( \Delta_c) C_{100} + ( g_{mc}) C_{010}, \\
	%&\quad + (-i  E) C_{110} + ( \Omega \mu e^{-i \phi} \sqrt{2}) C_{200}, \\
	0 &= (i  E) C_{000} + ( g_{mc}) C_{100} + ( \Delta_m) C_{010}, \\
	%&\quad + ( \Omega \mu e^{-i \phi}) C_{110} + ( g_{md}) C_{011} + (-i  E \sqrt{2}) C_{020}, \\
	0 &=  \Delta_d C_{001}, \\
	%+  \Omega \mu e^{-i \phi} C_{101} + (-i  E) C_{011}, \\
	0 &= (i  E) C_{100} + ( \Omega \mu e^{i \phi}) C_{010} + ( \Delta_c) C_{110} \\
	&\quad + ( \Delta_m) C_{110} +  g_{mc} \sqrt{2} \, C_{200} +  g_{mc} \sqrt{2} \, C_{020}, \\
	0 &=  \Omega e^{i \phi} C_{001} + ( \Delta_c) C_{101} + ( \Delta_d) C_{101} + ( g_{mc}) C_{011}, \\
	0 &= ( g_{md}) C_{010} + i  E C_{001} + ( g_{mc}) C_{101} + ( \Delta_m) C_{011} \\
	&\quad + ( \Delta_d) C_{011}, \\
	0 &=  \Omega \mu e^{i \phi} \sqrt{2} \, C_{100} +  g_{mc} \sqrt{2} \, C_{110} + (2  \Delta_c) C_{200}, \\
	0 &= i  E \sqrt{2} \, C_{010} +  g_{mc} \sqrt{2} \, C_{110} + (2  \Delta_m) C_{020}, \\
	0 &= 2  \Delta_d C_{002}. 
	\end{split}
	\end{equation}
	if We solve the steady state system of equations (\ref{eq:steady_state}), then each coefficient will be obtained in terms of the system's variables. 
	
	Solving $C_{200}=0$, the value for $E$ will be: 
	\begin{equation}
	E = -\frac{i \Omega \, \mu \, \Delta_m \, e^{i \phi} }{g_{mc}}
	\label{c200_E}
	\end{equation}
	By considering $ \Delta_m = \Delta_r - i \Delta_i$, equation (\ref{c200_E})  can be simplified to 
	\begin{equation}
	E = \frac{\Omega \mu}{g_{mc}}((\Delta_r sin\phi - \Delta_i cos\phi) - i(\Delta_r cos\phi + \Delta_i sin\phi))
	\end{equation}
	According to equations (\ref{H_dm}) and (\ref{H_dc}),  $\Omega\mu$ and $E$ are real values, so the parameter $E$ must satisfy the following constraint:
	\begin{equation}
	tan(\phi) = \frac{-\Delta_r}{\Delta_i}
	\label{c200_phi}
	\end{equation}
	Without loss of generality, we can choose $\phi$ such that $E$ becomes a positive value quantity. 
	
	The second order correlation function of photon is introduced as: 
	\begin{equation}
	g^{(2)}(0) = \frac{<c^\dagger c^\dagger c c>}{<c^\dagger c>^2} = \frac{2 |C_{200}|^2}{(|C_{100}|^2 + |C_{110}|^2 + |C_{101}|^2 + 2 |C_{200}|^2)^2}
	\label{photon_corr}
	\end{equation}
	
	According to (\ref{c200_phi}) and  (\ref{c200_E}), at each frequency in steady state situation, there are values for $\phi$ and $E$ to make $g^{(2)}(0)$ equal to zero. 
	
\begin{figure}[h!]
		\centering
		\includegraphics[scale=0.8]{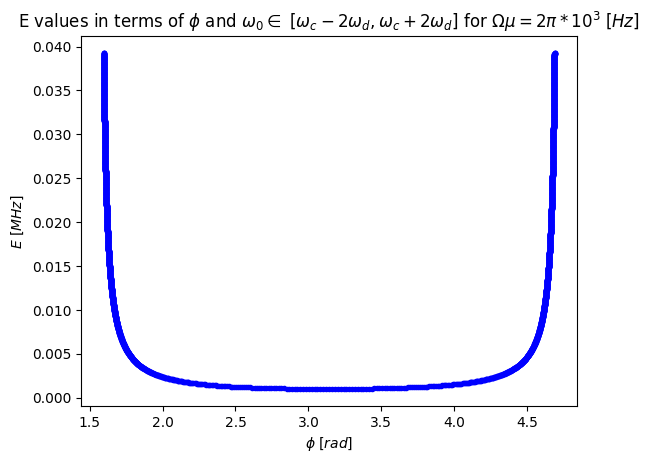} 
		\caption{E values in terms of $\phi$. } 
		\label{Evalues.png}
	\end{figure}

	To investigate how $E$ values change with respect to $\phi$, we will consider a simple numerical example. The system parameters are defined in table \ref{tab:system_parameters}, while $\omega_0$ varies in the range $[\omega_c-2\omega_d, \omega_c+2\omega_d]$ and $\Omega \mu = 2\pi \cross 10^3$. For each pump frequency, we obtain $\phi$ from equation (\ref{c200_phi}) and then calculate $E$ using (\ref{c200_E}). The values of $E$ for each $\phi$ are shown in Fig. \ref{Evalues.png}. This figure helps us to choose correct values for the phase $\phi$. 
	It is worth noting that the drive field can be taken weak enough when compared with the magnon (photon) decay rate $\kappa_m (\kappa_c)$.

	\begin{table}[h!]
		\centering
		\begin{tabular}{|l|l|l|}
			\hline
			\textbf{Variable} & \textbf{Value} \\
			\hline
			$\omega_c/ 2\pi= \omega_m/ 2\pi$ & $ 10 \, \text{GHz}$ \\
			\hline
			$\omega_d/ 2\pi$ & $10 \, \text{MHz}$ \\
			\hline
			$\kappa_c/ 2\pi = \kappa_m/ 2\pi$ & $ 1 \, \text{MHz}$ \\
			\hline
			$\gamma_d/ 2\pi$ & $100 \, \text{Hz}$ \\
			\hline
			$g_{mc}/2\pi = G_{md}/2\pi$ & $3.2 \, \text{MHz}$ \\
			\hline
			$T$ & $10 \, \text{mK}$ \\
			\hline
		\end{tabular}
		\caption{System parameters.}
		\label{tab:system_parameters}
	\end{table}
	
	Now we are ready to find the best values of system parameters to create the deepest available photon blockade.
	In the next section, we try to move to realistic values for the system parameters and also read the value of the photon blockade by direct numerical calculations.

	\section*{Results}
	
	Time evolution of the system's wave function is calculated by solving the set of differential equations (\ref{eq:exact}) for 10 wave function coefficients ($C_{ijk}$), which are complex variables as functions of time. Initial condition at $t=0$ is taken as $C_{000}=1$ and other coefficients equal to zero. System parameters have been taken from the table \ref{tab:system_parameters} for the best photon blockade situation. The frequency of the drive field ($\omega_d$) has been numerically scanned into the range $\omega_c - 2\omega_p$ up to $\omega_c + 3\omega_p$.
	Time evolution of the system has been calculated for sufficiently large time intervals in order to find the steady situation of the system for taken values of the system parameters. In this section we use the Radau method to find the exact solutions.
	The Radau method is a class of implicit Runge-Kutta methods, particularly suited for solving stiff ordinary differential equations (ODEs) and differential-algebraic equations (DAEs). It belongs to the family of collocation methods where the solution is approximated using collocation points at specific locations within the interval. Specifically, Radau IIA methods utilize collocation points that include the right endpoint of the interval, ensuring high stability properties. According to \cite{mate2023mathematical}, the stiffness ratio for the coefficient matrix of the equation (\ref{eq:exact}) is big enough to consider this problem an stiff one.
	
\begin{figure}[h!]
		\centering
		\includegraphics[scale=0.8]{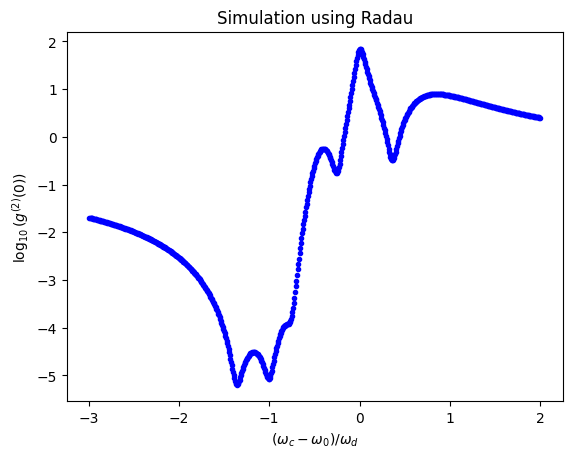} 
		\caption{$log(g^{(2)}(0))$ for different values of $\omega_0$.} 
		\label{simpleg2.png}
	\end{figure}

	Fig. \ref{simpleg2.png} demonstrates the $g^{(2)}(0)$ for different values of scaled detuning $\frac{\omega_c - \omega_w}{\omega_d}$. This figure clearly shows that drive fields with smaller frequencies than the cavity frequency are better choices for deeper photon blockade. Indeed, phase controlling of the system for finding a destructive superposition in the two-photon creation state is available with negative detuning.

\begin{figure}[h!]
		\centering
		\includegraphics[scale=0.8]{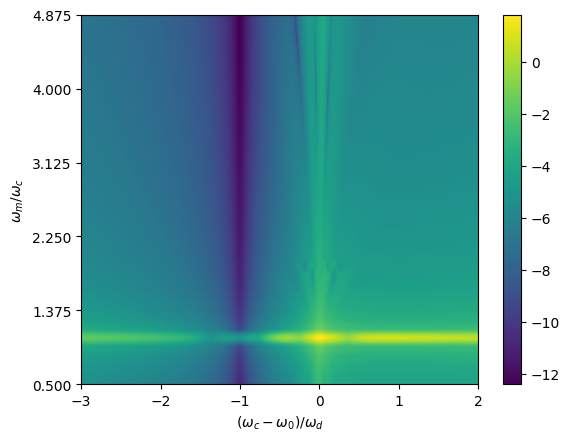} 
		\caption{$log(g^{(2)}(0))$ for different values of $\omega_0$ and $\omega_m$ with the system described in Fig. \ref{systemoutline.png}.} 
		\label{variableomegam.png}
	\end{figure}

	Another important parameter of the system is the magnon frequency $\omega_m$. The dependence of $g^{(2)}(0)$ on this parameter can also be plotted as in Fig. \ref{variableomegam.png}. The second-order correlation ($g^{(2)}(0)$) is a complicated function of adjustable parameters of the system. Thus, analyzing the system behavior through a two-parametric surface curve usually provides a better understanding of the system. Fig. \ref{variableomegam.png} demonstrates $g^{(2)}(0)$ as a function of magnon and drive field frequencies. For a better visualization, we have used the relative frequency for the magnon modes as $\left( \frac{\omega_m}{\omega_c} \right)$.
	Fig. \ref{variableomegam.png} has been plotted within the best choices for the system parameters so that $g^{(2)}(0)$ finds its lowest value. In this figure we have assumed $\omega_c$ to have the constant value mentioned in table. \ref{tab:system_parameters}. The $\frac{\omega_m}{\omega_c}$ plays a crucial  role in achieving more negative values across the entire range of the varying drive frequency. However, the deepest possible value occurs in the left half. So  the detuning of the drive frequency has the greatest effect on the depth of the $g^{(2)}(0)$ function. This is due to the significant control effect of the feedback which is applied through the cavity part of the system.
	
	If a quantum system is subjected to instability due to the sensitivity of some physical parameters, this issue can be reduced by arranging effective feedback. It is also possible to transfer control of the system to a part of it where experimental facilities for more precise adjustment of the system parameters are available and/or easier.
	
\begin{figure}[h!]
		\centering
		\includegraphics[scale=0.8]{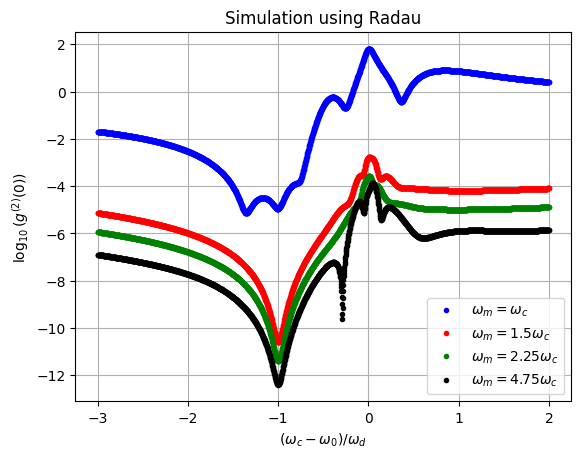} 
		\caption{Comparison of $log(g^{(2)}(0))$ for different values of $\omega_m$.} 
		\label{Randaucomparison.png}
	\end{figure}

	If we look at some specific points of Fig. \ref{variableomegam.png}, we will find that the system has a potential to reach blockade values lower than  $10^{-12}$ as shown in Fig. \ref{Randaucomparison.png}. It is evident that for values of $\omega_m > 2 \omega_c$, obtained values of $log(g^{(2)}(0))$ do not change significantly. 
	\begin{figure}[h!]
		\centering
		\includegraphics[scale=0.8]{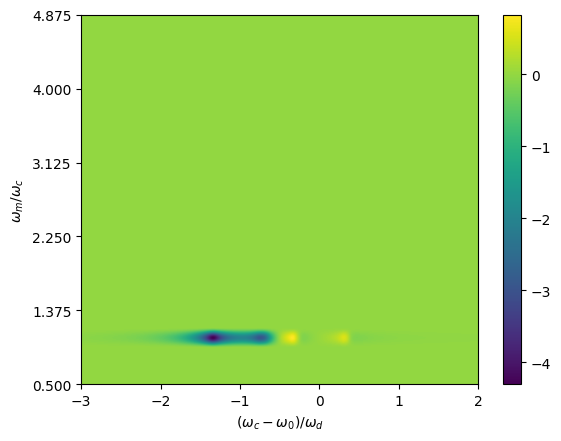} 
		\caption{$log(g^{(2)}(0))$ for different values of $\omega_0$ and $\omega_m$ without the feedback.} 
		\label{Nofeedback.png}
	\end{figure}
	
	The feedback loop plays an important role in achieving such deep values for the $log(g^{(2)}(0))$. To observe this, if we put $\Omega = 0$ and choose the value for magnon excitation $E = 2 \pi \cross 10^5$ (which is one order of magnitude less than $\kappa_m$), we will get Fig. \ref{Nofeedback.png}. In this figure, if the detuning is small ($\omega_m \approx \omega_c$) the blockade reaches its minimum values. So it is clear that only under the condition of adding a feedback loop, detuning between the magnon and photon frequencies can cause deep values of $log(g^{(2)}(0))$. This means that without a suitable phase feedback such deep photon blockade will not occur.

\begin{figure}[h!]
		\centering
		\includegraphics[scale=0.8]{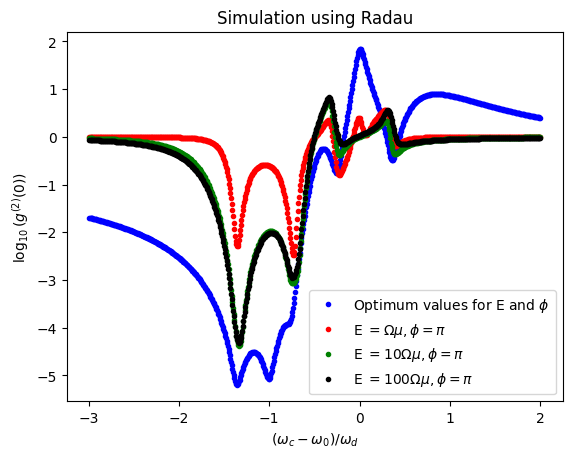} 
		\caption{$log(g^{(2)}(0))$ for different values of E and $\phi$.} 
		\label{OPtimuvalues.png}
	\end{figure}

	We should note that the equations (\ref{c200_phi}) and (\ref{c200_E}) are calculated under steady state conditions. However, these optimum values show perfect outcomes in comparison to constant phase and excitation. For instance, if we keep all the values the same and plot the blockade with both optimum values and constant values of E and $\phi$, we will get Fig. \ref{OPtimuvalues.png}. This figure clearly shows how phase controlled excitation can enhance results, that even with high excitations we can not achieve the same performance. 
	
	\section*{Conclusions}
	The photon blockade for a cavity magnomechanical system with a drive field feedback has been studied. We introduced optimum values for the phase and magnon frequency  in terms of other system parameters, which were extracted in the steady state situation of the system.
	We have shown that these issues can lead to much better performance when compared with a similar system without such phase feedback. 
	We also considered how a detuning between photon and magnon frequencies can enhance the values of photon blockade. This enables us to achieve the second order correlation function of the photon excitation ($g^{(2)}(0)$) values as deep as $10^{-12}$. 
	
	It was shown that the feedback procedure enables us to
	control the system behavior toward needed situations.
	Numerical simulations of the system indicate that the system
	is more stable against the sensitivity of system parameters, 
	as we have found for the magnon deturning in our studied
	model.
	
	Still there are several open questions which should be investigated.
	An important question is about preparing such a procedure for blockade in the magnonic part of the system.
	Arranging a phase controlled feedback
	acting on the mechanical or magnonic part of the system needs more attention.

	The feasibility of realizing a suitable phase feedback
	procedure for simultaneous control of different parts of a
	quantum system is a challenging issue, that requires deeper
	theoretical studies.
	We hope to delve into some of the above mentioned problems in further works.
	
	\bibliographystyle{plain}
	\bibliography{sample}

	\section*{Author contributions statement}
	Both authors contributed equally to all parts of this research.

\end{document}